# Development of Integration-Type Silicon-On-Insulator Monolithic Pixel Detectors by Using a Float Zone Silicon

S. Mitsui [a*], Y. Arai [b], T. Miyoshi [b], A. Takeda [c]

[a] *Venture Business Laboratory, Organization of Frontier Science and Innovation, Kanazawa University, Kakumamachi, Kanazawa, Ishikawa, 920-1192, Japan*
[b] *Institute of Particle and Nuclear Studies, High Energy Accelerator Research Organization (KEK), 1-1 Oho, Tsukuba, Ibaraki, 305-0801, Japan*
[c] *Faculty of Engineering, University of Miyazaki, 1-1 Gakuen kibanadai nishi, Miyazaki, Miyazaki, 889-2192, Japan*

**Abstract**

In this paper, we describe the development of monolithic pixel detectors by using a Silicon-on-Insulator (SOI) technology for X-ray and charged particle applications. The detectors are based on a 0.2 µm CMOS fully depleted SOI process (Lapis Semiconductor Co., Ltd). The SOI wafer consists of a thick high-resistivity substrate for sensing and a thin low resistivity Si layer for CMOS circuits.

We developed the integration-type SOI pixel detector, INTPIX4 mainly for X-ray imaging; it is made of a Float Zone (FZ) or Czochralski (Cz) silicon wafer. Since 2005, Cz SOI detectors were used initially. After 2011, FZ SOI detectors were successfully fabricated. In this paper, we state recent progresses and test results of the SOI monolithic pixel detector using a FZ silicon and compare them with the results obtained using the Cz detector.

## 1. Silicon-On-Insulator technology

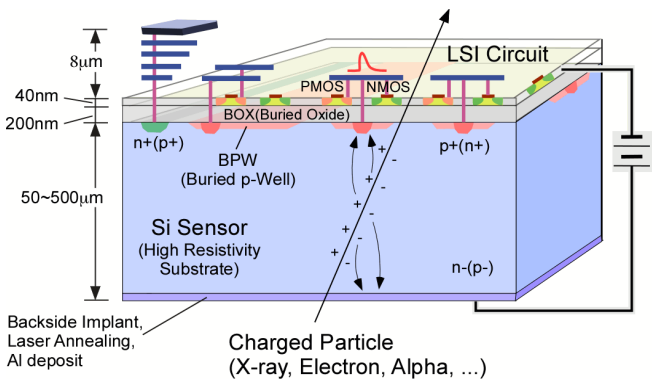

Fig. 1. Structures of the SOI pixel detector

We are developing SOI monolithic pixel detectors for X-ray and charged particle applications [Ref. 1-8]. The SOI detector is developed by bonding a thick high-resistivity wafer for sensing and a thin low-resistivity wafer for CMOS circuits [Fig. 1]. The SOI detector can have a high circuit density and reduce amounts of materials due to a monolithic detector. As the detector is developed by standard CMOS technology, complex functions can be implemented in a pixel. The detector does not comprise a mechanical bump bonding, thus ensuring high yield and low cost. A fully depleted detector has low capacitance at the sense node, that is, approximately 10 fF/pixel; thus, the conversion gain is high and the noise is suppressed to a low level. The SOI technology is based on an industrial standard technology, thus having a cost benefit and scalability. In addition, the SOI detectors have more advantages: no latch up and prevention of a single event effect. The detectors can run at low power, operate at a wide temperature range from 4 to 570 K. To prevent the back-gate effect, the buried p-well process is applied in each pixel.

The SOI detectors are fabricated using the 0.2 µm low-leakage fully depleted (FD)-SOI CMOS process of Lapis Semiconductor Co., Ltd. [Ref. 9]. Table 1 summarizes the main specifications of the SOI processes of Float zone (FZ) and Czochralski (Cz) silicon wafers.

SOI pixel detectors and test element groups (TEGs) are fabricated through the multiproject wafer (MPW)-runs organized by KEK approximately twice a year. A mask of 24.6 × 30.8 mm$^2$ is shared to reduce the production cost. Many Japanese and global institutes and universities contribute to the MPW run.

Table 1. Main specifications of the SOI processes of FZ and Cz silicon wafers

| Process | 0.2 µm low-leakage fully depleted (FD) SOI CMOS<br>1 Poly, 5 Metal layers (MIM Capacitor and DMOS option)<br>Core (I/O) voltage : 1.8 (3.3) V |
|---|---|
| SOI wafer<br>8 inch | Top Si : Cz, ~18 Ω-cm, p-type, ~ 40 nm thick<br>Buried Oxide: 200 nm thick<br>Handle wafer thickness: 725 µm<br>-> thinned up to ~300 µm in Lapis' processes<br>-> thinned up to ~50 µm in commercial processes<br>Handle wafer: Cz (N) ~700 Ω-cm, FZ (N) >3 kΩ-cm |
| Back side process | Mechanical Grind → Chemical Etching → Back side Implant → Laser Annealing → Al plating |

----
*corresponding author.
E-mail address: smitsui@staff.kanazawa-u.ac.jp (S. Mitsui)

## 2. INTPIX 4

INTPIX4 is an integration-type SOI pixel detector mainly used for X-ray imaging. Its chip size is 10.3 × 15.5 mm$^2$, and its effective area is 8.704 × 14.144 mm$^2$. In addition, the number of pixels is 512 × 832, each of size 17 × 17 µm$^2$ and comprising a correlated double sampling (CDS) circuit to reduce the reset noise. Fig. 2 shows the schematic of the on-pixel circuit. The detector has 13 parallel output lines. The resistivity and thickness of the N-type wafer are respectively over 700 Ω-cm and 260 µm for Cz, and over 7 kΩ-cm and 500 µm for FZ. Both the front- and back-side illuminations are available.

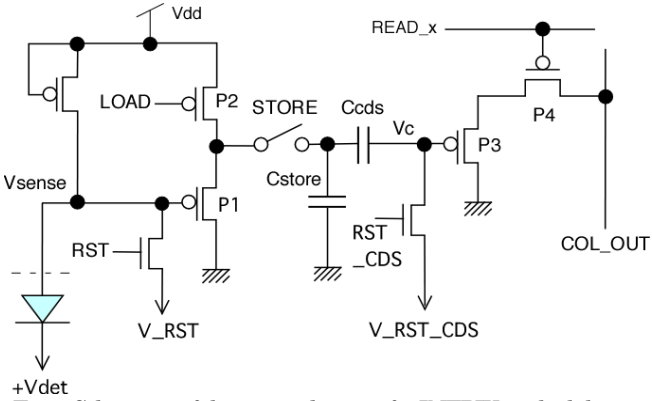

Fig.2. Schematic of the on-pixel circuit for INTPIX4, which has a CDS function and 2 capacitors: Cstore and Ccds.

## 3. Readout board

Fig. 3 shows a general-purpose evaluation board of SOI pixel detectors, that is, SOI evaluation board with Sitcp 2 (SEABAS2), and a sub board for INTPIX4. The SEABAS2 has Gigabit Ethernet for data transfer connected to a data acquisition PC. The user field-programmable gate array (FPGA) vertex5 and the FPGA for SiTCP [Ref. 10] are mounted on the board. The operation clock is of 50 MHz with an operation voltage of + 3.3 V, and −3.3 V for nuclear instrument module (NIM) signals. We implemented a 12 bit 16ch ADC, a 4ch DAC, and an NIM I/O interface. Owing to the Gigabit Ethernet, the SEABAS2 can read rapidly, it utilizes approximately 90 fps with 1 chip of INTPIX4. Thus, even an X-ray movie can be captured.

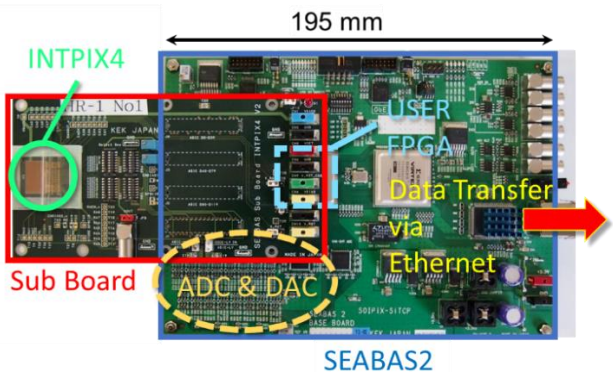

Fig. 3. SEABAS2 and INTPIX4 sub board

Fig. 4 shows the X-ray image of a medical stent wire; this was shot by using a 33.3 keV monochromatic X-ray at a photon factory in KEK. Its integration time is 200 µs; 250 frames are captured, thus making the total integration time as 50 ms. The stent wire is 40 µm in diameter and made of nickel titanium. The amount of acrylic case is consistent with the human body. The stent wire is clearly seen by using INTPIX4 and SEABAS2.

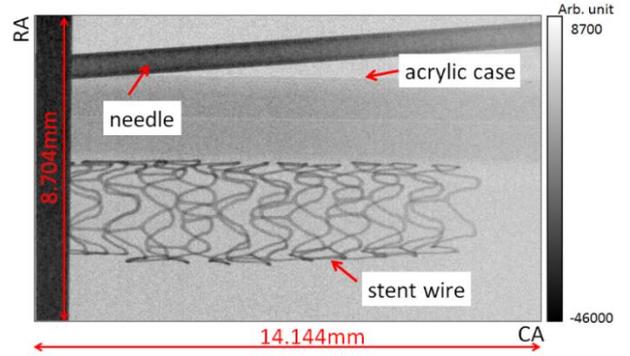

Fig. 4. X-ray image of medical stent wire.

## 4. Performance of FZ and Cz SOI detectors

### 4.1 Leakage current improvement

Initially, Cz SOI detectors were used since 2005. After 2011, FZ SOI detectors were successfully fabricated and the leakage current decreased [Fig. 5]. In this paper, the performance of FZ INTPIX4 detectors is principally reported, along with the comparison of FZ and Cz SOI detectors.

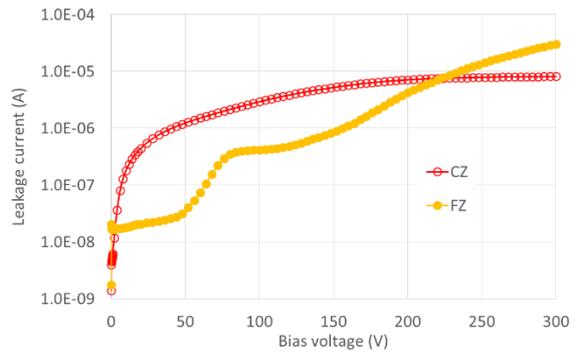

Fig. 5. Comparison of leakage current in FZ and Cz detectors. The leakage current is reduced to under 200 V for Fz detectors.

### 4.2 Spatial resolution

We evaluated the spatial resolution through direct radiography. The X-ray is illuminated from back-side by using a test chart with 12.5, 16, and 20 lp/mm (line pairs par mm) slits with window widths of 40, 31.25, and 25 µm, respectively. Fig. 6 shows a negative photograph of the X-ray image of the FZ detector and its gray values. The power of the X-ray generator is 30 kV, 20 mA, in which a Mo target is installed, and a characteristic X-ray of 17.48 keV is generated. In addition, the applied bias voltage of the FZ detector is 100 V, with 250 μs integration time; we integrated 1000 frames.

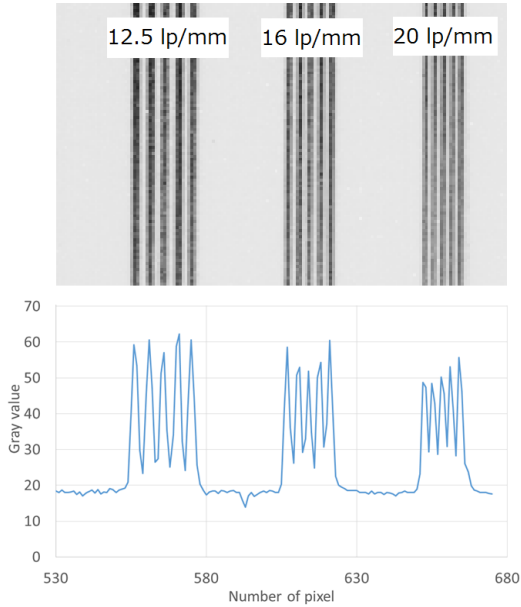

Fig. 6. The upper portion is a negative photograph of direct radiography, and the lower graph are gray values of 12.5, 16, and 20 lp/mm.

The contrast transfer function (CTF) is defined as follows:
$$\text{CTF}(\%) = \frac{I_{MAX} - I_{GAP}}{I_{MAX} - I_{MIN}}$$
where $I_{MAX}$ is the average of gray values at slits, $I_{GAP}$ is the average of gray values between slits, and $I_{MIN}$ is the average of gray values of base line not irradiated by X-rays. Fig. 7 shows the CTF of the 500 μm thick FZ detector and 260 μm thick Cz detector [Ref. 11]. The CTFs of FZ and Cz are respectively 66% and 78% at 20 lp/mm. The CTF of FZ is slightly worse than that of Cz because of its thickness. The generated electron-hole pairs spread while drifting from the back side in a thicker Si bulk.

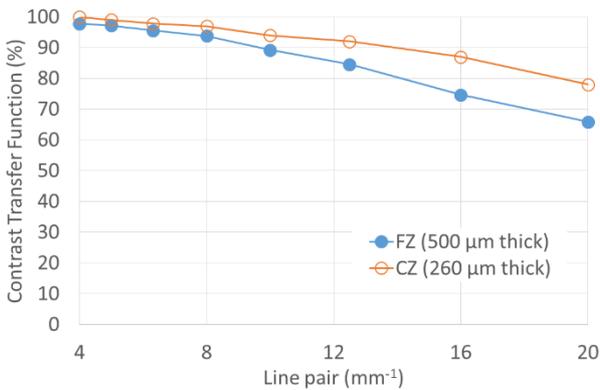

Fig.7. CTF of the FZ detector of 500 μm thickness and the Cz detector of 260 μm thickness.

### 4.3 Circuit gain

The circuit gain in a pixel is evaluated using the coefficient of proportionality of the analog voltage output to the CDS reset voltage. The analog output is proportional to the input CDS reset voltage over the threshold voltage, up to the saturation voltage [Fig. 8]. Furthermore, the circuit gain is obtained through the slope of the proportional region. Fig. 9 shows the circuit gain of all the pixels. Moreover, the FZ detector resulted in a circuit gain of $0.87 \pm 0.02$, which is consistent with the circuit gain of Cz, that is, $0.823 \pm 0.001$ [Ref. 11].

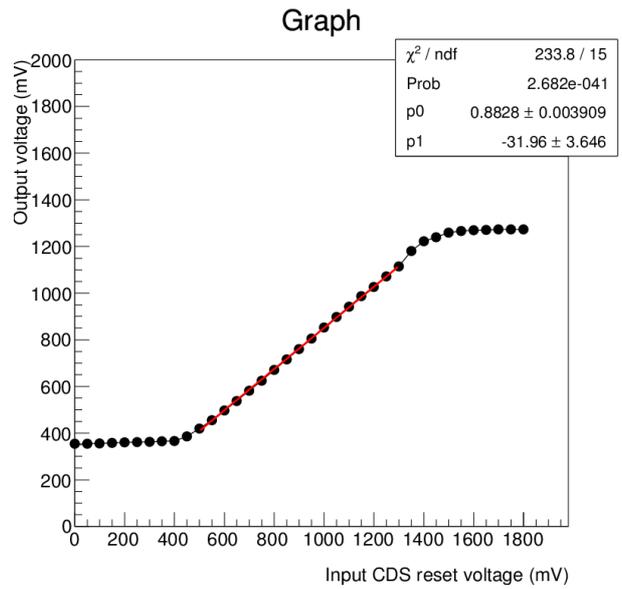

Fig. 8. Dependence of the analog output voltage on the CDS reset voltage.

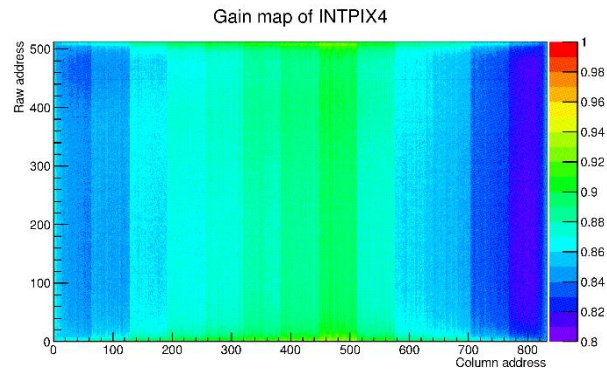

Fig. 9. Uniformity of circuit gain of FZ INTPIX4.

### 4.4 Energy spectrum

We obtained the energy spectrum of characteristic X-rays of Mo and Cu targets by using the X-ray generator. The X-rays are irradiated from behind the detector, and the spectrum is captured at room temperature. The bias voltage is 200 V. The power of X-ray tube with a Mo target is 30 kV, 20 mA and with Cu target is 20 kV, 20mA. The collected charge is evaluated by clustering some pixels because the electron-hole pairs generated through X-rays spread in the Si bulk. The collected charges of the Mo and Cu targets are evaluated by clustering 9 and 25 pixels, respectively. The energies of the characteristic X-rays of Mo-$K_\alpha$, Mo-$K_\beta$, and Cu-$K_\alpha$ are 17.48, 19.61, and 8.05 keV, respectively. The resulting detector gain is 13.1 µV/e, which is calculated through the slope of the output voltage at each energy [Fig. 10]. The full-width half-maximum (FWHM) of 624 electrons is obtained as 12.8 % at 17.48 keV. The detector gain and FWHM of the Cz detector are 12.61µV/e and 13.4% at 13.95 keV [Ref. 11]. The detector gain and FWHM of both FZ and Cz are consistent.

4.5 I–V characteristics

We investigated the I–V characteristics at temperatures from −60 °C to +20 °C at 10 °C steps. At each temperature, there is nosignificant breakdown of up to 500 V for the FZ detector [Fig. 11]. The leakage current increase of approximately 280 V is probably caused by the depletion region reaching to back surface. The reason for the increase of approximately 50 V is under investigation. The activation energy can be obtained through the Arrhenius plot as follows:

$$I = Ae^{\frac{-Ea}{kT}},$$

where $I$ is a leakage current, $A$ is a coefficient, $E_a$ is an activation energy, $k$ is the Boltzmann constant, and $T$ is the absolute temperature. After plotting the current value at each temperature and fitting the plots with the Arrhenius equation, the activation energy is obtained as the slope of the curves [Fig. 12], and is 0.55 eV at 30 V. It is consistent with that of the Cz detector: 0.56 eV [Ref. 11]. The activation energy of the FZ detector is evaluated at 30 V because of the increase in the leakage current over 50 V. Moreover, the leakage current at low temperature region is avoided in the calculation because it exceeds the measurement limit of an ammeter.

4.6 C–V characteristics

We measured the body capacitance of FZ and Cz detectors. The body capacitance is measured using an LCR meter by applying a bias voltage of up to 400 V at room temperature. The inverse square of the body capacitance is proportional to the bias voltage until the detector is FD as follows:

$$\frac{1}{C_{bulk}^2} = \frac{2}{S^2 q\epsilon} \frac{N_A + N_D}{N_A N_D} V,$$

where $C_{bulk}$ is the body capacitance of the detector, $S$ is the depleted area, $q$ is the elementary charge, $\epsilon$ is the permittivity of silicon, $N_A$ is the bulk acceptor density, $N_D$ is the bulk donor density, and $V$ is an applied voltage. After complete depletion, the body capacitance becomes constant. The capacitance is measured at the measurement frequency of the LCR meter from 1 kHz to 1 MHz [Fig. 13 and 14]. The complete depletion voltages of FZ are evaluated as 321 V at 100 kHz and 325 V at 300 kHz. Furthermore, those of Cz are evaluated as 238 V at 30 kHz and 239 V at 10 kHz.

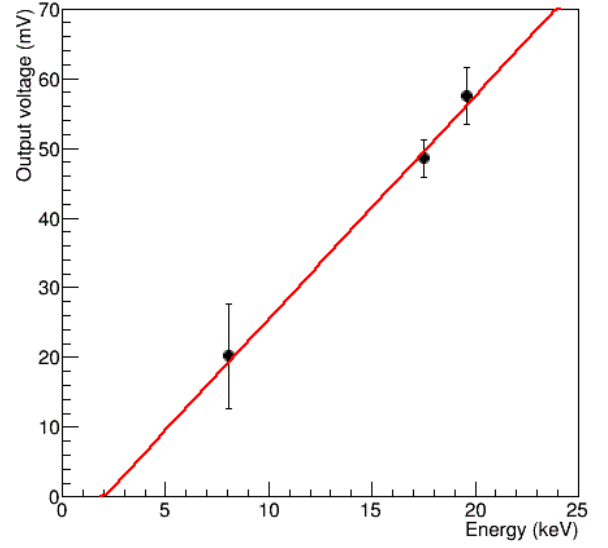

Fig. 10. Detector gain of FZ SOI detector.

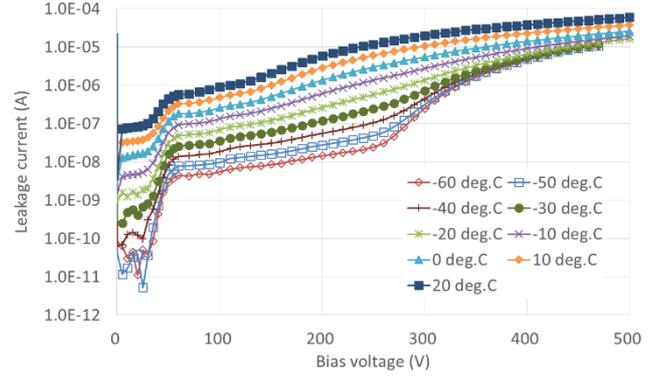

Fig. 11. I–V characteristics of FZ detector from −60 °C to +20 °C for up to 500 V.

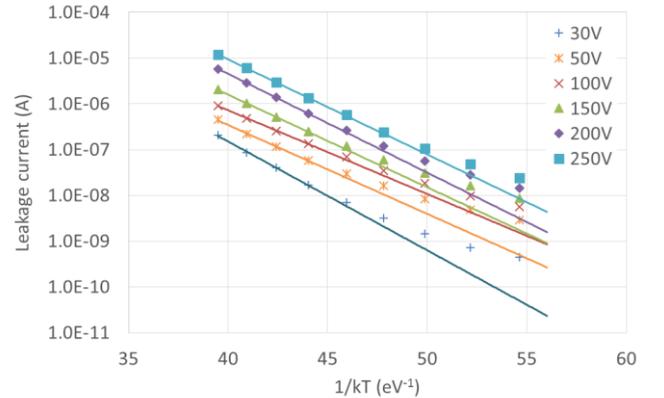

Fig. 12. Arrhenius plots from 30 to 250 V of bias voltage.

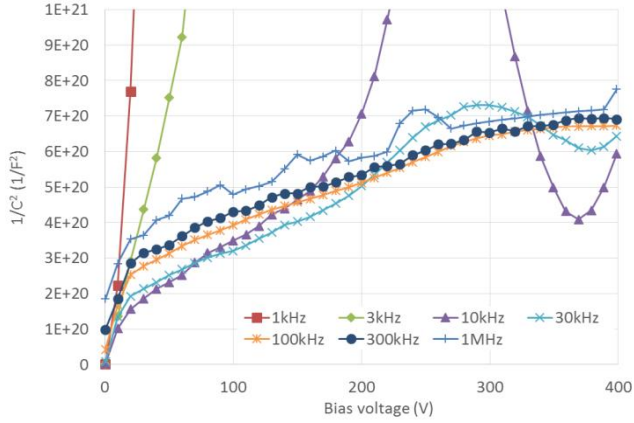

Fig. 13. Bias voltage dependence of 1/(the body capacitance)$^2$ with the FZ detector from 1 kHz to 1 MHz of measurement frequency

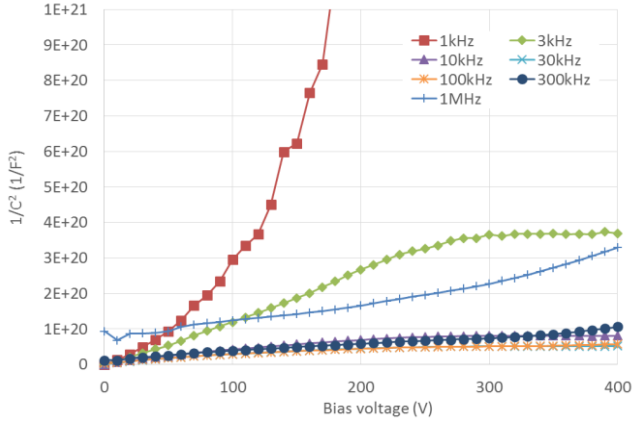

Fig. 14. Bias voltage dependence of 1/(the body capacitance)$^2$ with the Cz detector from 1 kHz to 1 MHz of measurement frequency

4.7 Beta-ray source test

We performed beta-ray source tests of $^{90}$Sr. The beta rays from $^{90}$Sr source include 2.28 MeV beta rays of $^{90}$Y, which can be assumed as minimum ionization particles (MIPs). The collimator has a hole of 3 mm diameter. The Al attenuator is set on the collimator to reduce low-energy beta rays. The trigger scintillator has dimensions of $10 \times 10 \times 10$ mm$^3$. Moreover, the thicknesses of FZ and Cz detectors are 500 and 260 μm respectively, with corresponding applied bias voltages of 200 and 100 V, and integration times of 1000 and 400 μs. Fig. 14 and 15 show the energy distribution of FZ and Cz detectors, for which the analog digital units of the signal peaks are 374 and 235, pedestal sigma are approximately 5 and 13, and the signal-to-noise ratios are approximately 75 and 18, respectively. Thus, MIP is observable by using the SOI detector.

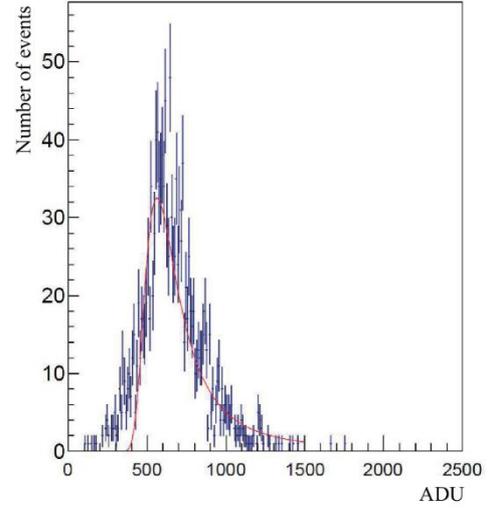

Fig. 15. Energy distribution of beta rays of $^{90}$Sr by using the FZ detector

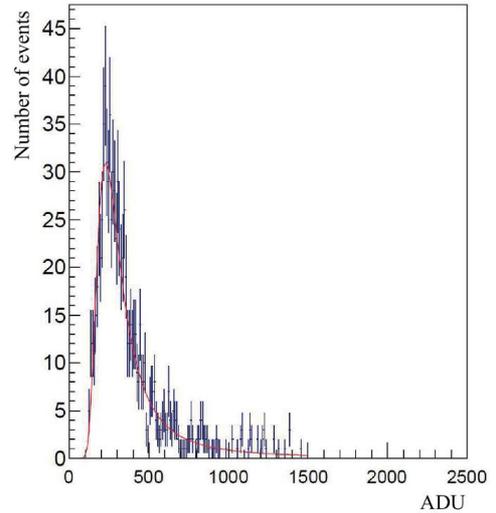

Fig. 16. Energy distribution of beta rays of $^{90}$Sr by using the Cz detector.

## 5. Summary

We are developing monolithic pixel detectors by using SOI technology for X-ray and charged particle applications. We conclude that the integration-type SOI pixel detector INTPIX4 functions successfully, that is, it can capture X-ray images at high spatial resolutions, high contrast, and high speed. In addition, FZ detectors show a good performance similar to that of Cz. The leakage current in FZ detector is less than that in Cz. However, the fully depleted voltages of FZ are higher than of Cz because of its thicker bulk. In addition, the approximately 10 times higher resistivity of FZ detector enables thicker sensing depth at a moderate bias voltage. Further, the detector gain, activation energy, and energy resolution of FZ detector is consistent with those of Cz. Both Fz and Cz detectors can observe MIP with a good signal-to-noise ratio. They are valuable for particle-tracking after thinning.